\documentclass[dvips]{article}
\usepackage{supertabular,lscape,epsfig}
\usepackage{amssymb}
\usepackage{amsmath}

\usepackage[polish]{babel}
\usepackage[T1]{fontenc}
\usepackage[latin2]{inputenc}
\usepackage{pslatex}

\DeclareSymbolFont{ppa}{OT1}{ppl}{m}{it}
\DeclareMathSymbol{\vv}{\mathalpha}{ppa}{'166}

\begin{document}

\newcommand{\dd}{\,{\rm d}}
\newcommand{\ie}{{\it i.e.},\,}
\newcommand{\etal}{{\it et al.\ }}
\newcommand{\eg}{{\it e.g.},\,}
\newcommand{\cf}{{\it cf.\ }}
\newcommand{\vs}{{\it vs.\ }}
\newcommand{\zdot}{\makebox[0pt][l]{.}}
\newcommand{\up}[1]{\ifmmode^{\rm #1}\else$^{\rm #1}$\fi}
\newcommand{\dn}[1]{\ifmmode_{\rm #1}\else$_{\rm #1}$\fi}
\newcommand{\upd}{\up{d}}
\newcommand{\uph}{\up{h}}
\newcommand{\upm}{\up{m}}  
\newcommand{\ups}{\up{s}}
\newcommand{\arcd}{\ifmmode^{\circ}\else$^{\circ}$\fi}
\newcommand{\arcm}{\ifmmode{'}\else$'$\fi}
\newcommand{\arcs}{\ifmmode{''}\else$''$\fi}
\newcommand{\MS}{{\rm M}\ifmmode_{\odot}\else$_{\odot}$\fi}
\newcommand{\RS}{{\rm R}\ifmmode_{\odot}\else$_{\odot}$\fi}
\newcommand{\LS}{{\rm L}\ifmmode_{\odot}\else$_{\odot}$\fi}

\newcommand{\Abstract}[2]{{\footnotesize\begin{center}ABSTRACT\end{center}
\vspace{1mm}\par#1\par   
\noindent
{~}{\it #2}}}

\newcommand{\TabCap}[2]{\begin{center}\parbox[t]{#1}{\begin{center}
  \small {\spaceskip 2pt plus 1pt minus 1pt T a b l e}
  \refstepcounter{table}\thetable \\[2mm]
  \footnotesize #2 \end{center}}\end{center}}

\newcommand{\TableSep}[2]{\begin{table}[p]\vspace{#1}
\TabCap{#2}\end{table}}

\newcommand{\FigCap}[1]{\footnotesize\par\noindent Fig.\  %
  \refstepcounter{figure}\thefigure. #1\par}

\newcommand{\TableFont}{\footnotesize}
\newcommand{\TableFontIt}{\ttit}
\newcommand{\SetTableFont}[1]{\renewcommand{\TableFont}{#1}}

\newcommand{\MakeTable}[4]{\begin{table}[htb]\TabCap{#2}{#3}
  \begin{center} \TableFont \begin{tabular}{#1} #4
  \end{tabular}\end{center}\end{table}}

\newcommand{\MakeTableSep}[4]{\begin{table}[p]\TabCap{#2}{#3}
  \begin{center} \TableFont \begin{tabular}{#1} #4
  \end{tabular}\end{center}\end{table}}

\newenvironment{references}%
{
\footnotesize \frenchspacing
\renewcommand{\thesection}{}
\renewcommand{\in}{{\rm in }}
\renewcommand{\AA}{Astron.\ Astrophys.}
\newcommand{\AAS}{Astron.~Astrophys.~Suppl.~Ser.}
\newcommand{\ApJ}{Astrophys.\ J.}
\newcommand{\ApJS}{Astrophys.\ J.~Suppl.~Ser.}
\newcommand{\ApJL}{Astrophys.\ J.~Letters}
\newcommand{\AJ}{Astron.\ J.}
\newcommand{\IBVS}{IBVS}
\newcommand{\PASP}{P.A.S.P.}
\newcommand{\Acta}{Acta Astron.}
\newcommand{\MNRAS}{MNRAS}
\renewcommand{\and}{{\rm and }}
\section{{\rm REFERENCES}}
\sloppy \hyphenpenalty10000
\begin{list}{}{\leftmargin1cm\listparindent-1cm
\itemindent\listparindent\parsep0pt\itemsep0pt}}%
{\end{list}\vspace{2mm}}
 
\def\TYLDA{~}
\newlength{\DW}
\settowidth{\DW}{0}
\newcommand{\dw}{\hspace{\DW}}

\newcommand{\refitem}[5]{\item[]{#1} #2%
\def\REFARG{#3}\ifx\REFARG\TYLDA\else, {\it#3}\fi
\def\REFARG{#4}\ifx\REFARG\TYLDA\else, {\bf#4}\fi
\def\REFARG{#5}\ifx\REFARG\TYLDA\else, {#5}\fi.}

\newcommand{\Section}[1]{\section{#1}}
\newcommand{\Subsection}[1]{\subsection{#1}}
\newcommand{\Acknow}[1]{\par\vspace{5mm}{\bf Acknowledgements.} #1}
\pagestyle{myheadings}

\newfont{\bb}{ptmbi8t at 12pt}
\newcommand{\xrule}{\rule{0pt}{2.5ex}}  
\newcommand{\xxrule}{\rule[-1.8ex]{0pt}{4.5ex}}  
\def\thefootnote{\fnsymbol{footnote}}
\begin{center}

{\Large\bf XROM and RCOM:
\vskip5pt
Two New OGLE-III Real Time Data Analysis Systems}
\vskip1.5cm
{\bf A.~~ U~d~a~l~s~k~i}
\vskip5mm
{Warsaw University Observatory,
Al.~Ujazdowskie~4, 00-478~Warszawa, Poland\\
e-mail: udalski@astrouw.edu.pl}
\end{center}

\Abstract{We describe two new OGLE-III real time data analysis systems:
XROM and RCOM. The XROM system has been designed to provide continuous real
time photometric monitoring of the optical counterparts of X-ray sources
while RCOM system provides real time photometry of R~Coronae Borealis
variable stars located in the OGLE-III fields. Both systems can be used for
triggering follow-up observations in crucial phases of variability episodes
of monitored objects.}{Surveys -- Techniques: photometric -- X-rays: stars
-- Stars: AGB and post-AGB}

\Section{Introduction} 
The Optical Gravitational Lensing Experiment (OGLE) is a long term large
scale sky survey regularly monitoring the most dense stellar fields in
the sky (Udalski \etal 1992, Udalski, Kubiak and Szymański 1997, Udalski
2003). The OGLE project started originally in 1992 as a first generation
microlensing survey and contributed in its subsequent phases to many
fields of modern astrophysics like stellar astrophysics, extrasolar
planet searches, gravitational lensing and others. Huge databases of
photometric measurements of hundreds of millions stars spanning several
years provide a unique opportunity for data mining, performing
statistical analyzes of huge samples of particular objects or conducting
analysis of long term behavior of selected classes of stars.

One of the most important results of the OGLE survey was the implementation
of real time data analysis systems that allow monitoring of selected
variable objects in almost real time. Advantages of this approach are
obvious, especially in the case of transient or non-periodic variable
objects. For instance, one can carefully prepare different kind of
follow-up observations, knowing current photometric behavior of a selected
object. Microlensing field is the best example here. The OGLE project was
the first to implement the so called Early Warning System (EWS, Udalski
\etal 1994) -- the system that detected on-going microlensing events in
their early phases. Information on such events was made public. Several
microlensing follow-up teams were formed in the several past years to
observe intensively the already discovered ongoing microlensing
phenomena. This strategy turned out to be very successful leading to the
important discoveries like microlensing extrasolar planets (Udalski \etal
2005, Beaulieu \etal 2006, Gould \etal 2006, Gaudi \etal 2008) or Magellanic
Cloud microlensings (Afonso \etal 2000, Dong \etal 2007).

OGLE-III phase of the OGLE project, that started on June 12, 2001 and has
been conducted up to now, was a significant extension of the OGLE
survey. Much larger observing capabilities made it possible to cover
practically entire area of the LMC and SMC and large fraction of the
Galactic bulge. Also new data analysis systems were implemented during this
phase (Udalski 2003). Beside the EWS system allowing the discovery of about
600 microlensing events every year, two new systems were developed after
the first four seasons of the OGLE-III phase: EEWS and NOOS.

The EEWS system was designed to detect in the real time anomalies of
microlensing events from a single mass microlensing. The implementation of
this system became an important step -- thanks to it the vast majority of
non-standard microlensing events, including many planetary microlensings,
were detected in almost real time and the information was passed to other
microlensing groups. EEWS system also allowed to switch the OGLE observing
mode from the standard survey mode to follow up mode where the observations
of a particular object were done with much higher cadence -- dependent on
the variability rate.

The second real time system NOOS (Udalski 2003) was designed to detect in
real time the transient stellar objects that brighten strongly enough to be
seen in the OGLE images for some time. This class of objects include
supernovae (SNe), long term variable stars, microlensing of very faint
stars (non-detectable in the regular OGLE photometry range) etc. A few new
SNe were detected soon after the implementation of this system (Udalski
2004).

Finally, the real time monitoring of the Einstein Cross gravitational lens
(QSO 2237+0305) provides the real time photometry of four images of the
quasar. This object is one of the most important gravitational lenses and
unique OGLE photometric dataset (Woźniak \etal 2000, Udalski \etal 2006)
was often used for its modeling. Continuous monitoring of the quasar images
allows early detection of potential caustic crossings or cusp approaches in
this lens. These events are crucial for proper modeling and understanding
the gravitational lenses and provide an opportunity to estimate the quasar
size.

In this note we present two new real time OGLE-III data analysis systems
implemented recently: XROM and RCOM. They allow real time monitoring of
selected classes of highly variable optical objects. The photometry
provided by these systems is available to the astronomical community from
the OGLE Internet archive.

\Section{XROM: X-Ray Variables OGLE Monitoring System}
X-ray astronomy is one of the most rapidly developing branches of modern
astrophysics. New space missions provide more and more exciting data in
this wavelenght range and the data flow accelerates. Nevertheless in the
majority of cases the proper interpretation of observed X-ray behavior of
detected objects requires observations in other wavelengths as well,
including the optical range.

The dense OGLE-III fields like the Magellanic Clouds or Galactic bulge
include many X-ray sources. Part of them has been successfully identified
with the optical counterparts. For example, the SMC contains a large sample
of X-ray pulsars discovered during the past few years (Coe \etal
2005). OGLE data has already been used for interpretation of some of these
objects (Coe \etal 2005, McGowan \etal 2008).

Continuous optical monitoring of counterparts of X-ray sources is very
important, as many of them undergo large optical variations, eruptions
etc. likely related to the X-ray activity. Therefore, many planned X-ray
follow-up observations may be much better tuned-up when the current optical
state and behavior of these objects is known.
\begin{figure}[htb]
\centerline{\includegraphics[width=10cm, bb=25 50 505 430]{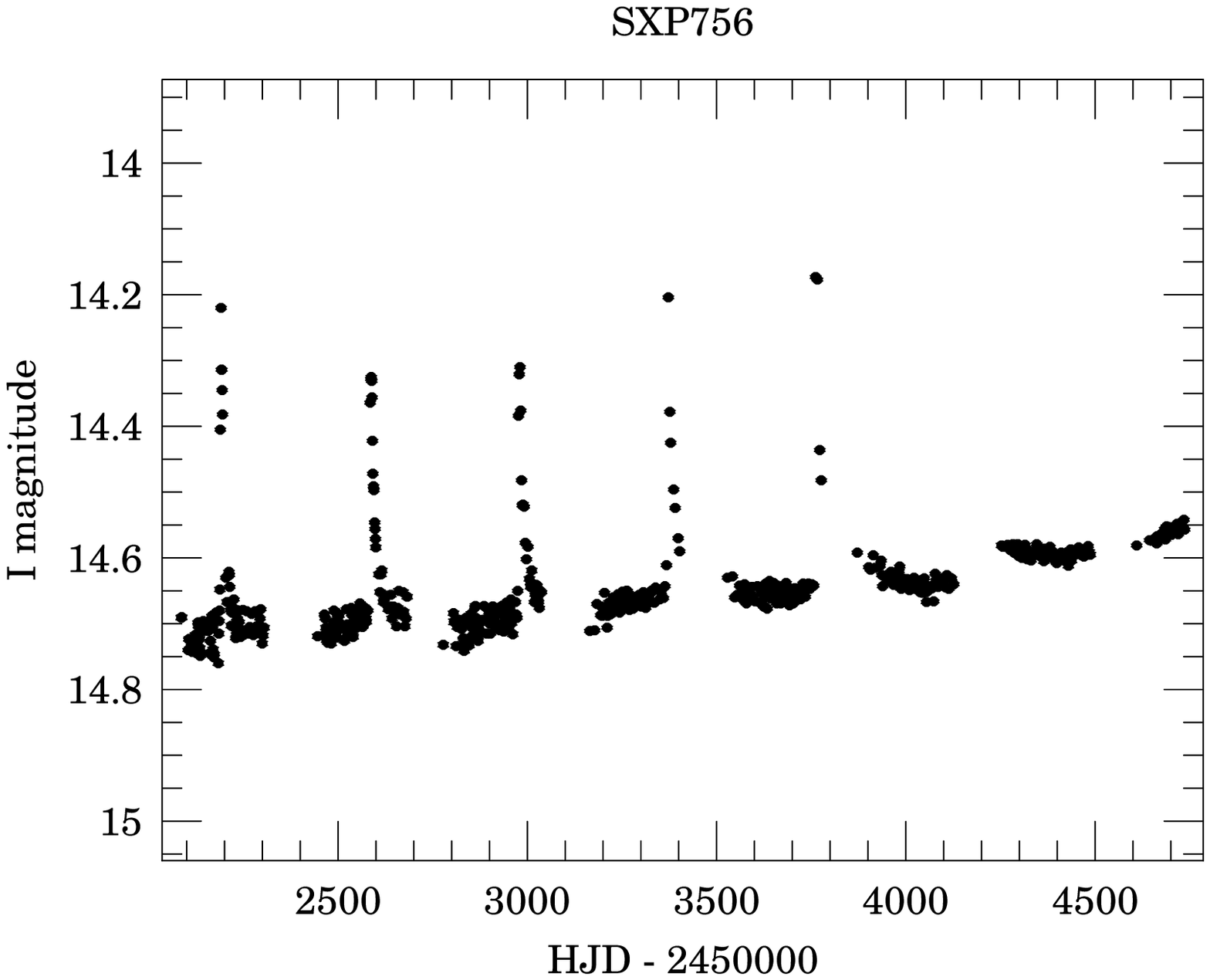}}
\FigCap{Light curve of the X-ray pulsar SXP~756.}
\end{figure}

The OGLE-III XROM system provides continuous photometric coverage of a
selected sample of known optical counterparts of X-ray sources located in
the OGLE-III fields. The initial sample contains 52 objects. It can be
easily extended with other or newly detected objects. Photometry of the
XROM objects is typically updated after each clear night. Fig.~1 presents
the OGLE-III light curve of one of such objects: SXP~756.

The interactive access to the XROM objects is provided {\it via} the main
OGLE WWW page:

\begin{center}  
{\it http://ogle.astrouw.edu.pl}\\
\end{center}

The structure of the page is similar to other OGLE real time system
pages. After selecting an object, the object page is invoked providing the
basic information: its OGLE identification, RA/DEC coordinates, finding
chart and two light curve plots: one showing the entire light curve and the
second one showing the last 60 days. The photometry is obtained through the
{\it I}-filter and it is only roughly calibrated with accuracy of the zero
points of $\pm0.1{-}0.2$~mag.

The photometry can be download from the OGLE archive:

\begin{center}  
{\it ftp://ftp.astrouw.edu.pl/ogle3/xrom}
\end{center}

\Section{RCOM: OGLE Real Time Monitoring of R~CrB Variable Stars}
R~CrB stars form a group of stars that reveal dramatic variability
episodes. Their brightness can fade by a few magnitudes in the time scale
of several days. These fading episodes are unpredictable and can last for
months. After that period the brightness of these stars gradually recovers
to the original state.
\begin{figure}[htb]
\centerline{\includegraphics[width=10.4cm, bb=25 50 505 440]{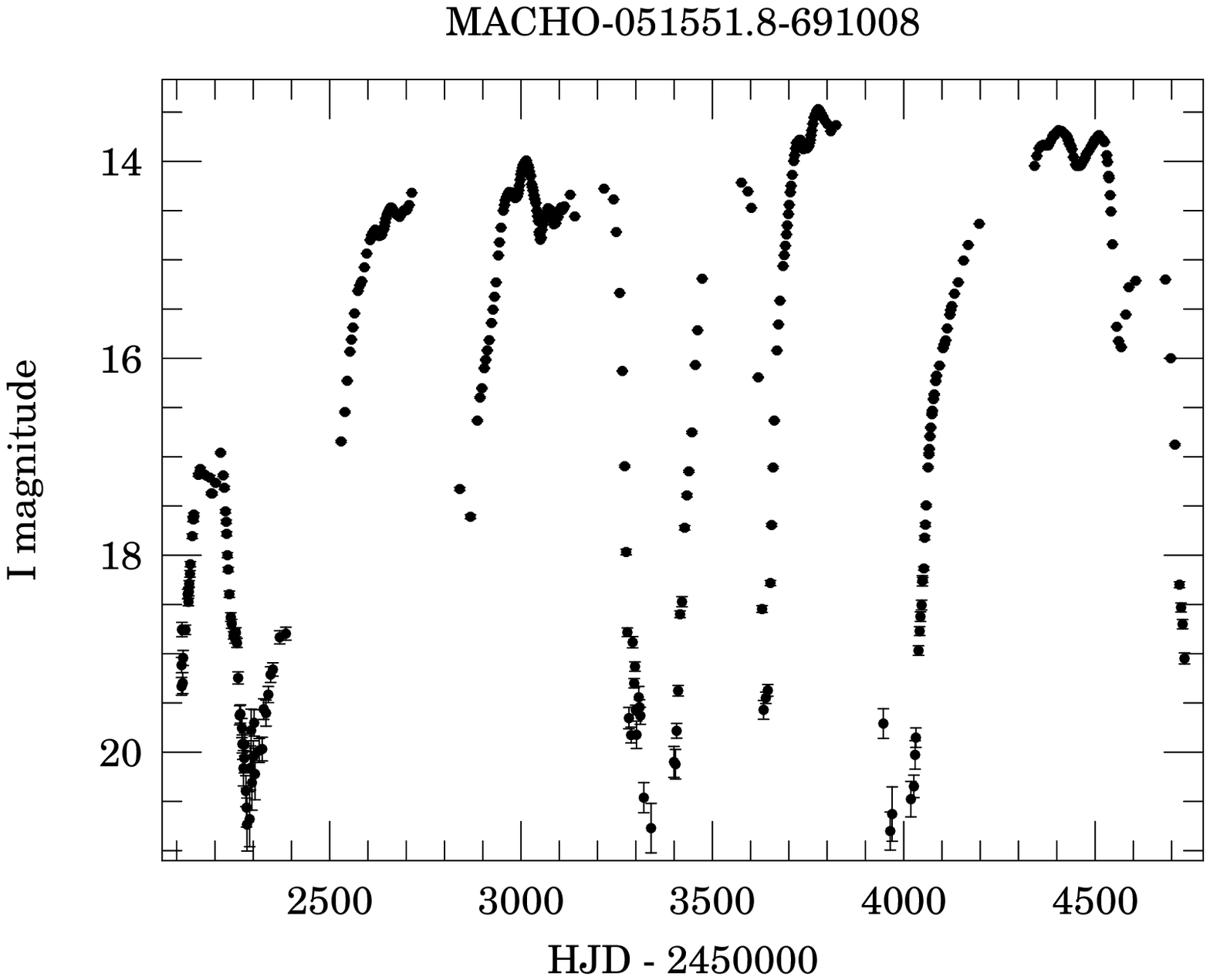}}
\FigCap{Light curve of the R~CrB type star, MACHO-051551.8-691008,
monitored by the OGLE-III CROM system.}
\end{figure}

It is believed that fading is related to the formation of dust clouds
over the surface of these stars. When they disperse the brightness returns
to the unobscured level. Some such clouds were directly observed. R~CrB
variables are very small group of stars -- only about 50 is known in the
Galaxy (Tisserand \etal 2008) and about 20 in the Magellanic Clouds (Alcock
\etal 2001, Tisserand \etal 2004).

R~CrB stars provide an opportunity of studying the late stages of stellar
evolution. To clarify their evolutionary status extensive follow-up
observations are needed in the most dramatic fading or rising phases of the
variability episodes. As the episodes are unpredictable only continuous
observations of R~CrB variables may trigger such follow-up programs.

OGLE-III fields are ideal for the real time data analysis system monitoring
R~CrB variables. They contain many R~CrB stars, both in the Magellanic
Clouds and the Galactic bulge. The RCOM system was designed to continuously
monitor a sample of R~CrB stars from the OGLE-III fields. Initially the
sample consists of 23 objects but it can be extended when the new objects
are detected. Unfortunately most of the known R~CrB stars in the Galactic
bulge are saturated in the OGLE-III reference images, therefore their
photometry is not available. Fig.~2 shows an example of the light curve of
one of the R~CrB objects monitored by the CROM system,
MACHO-051551.8-691008.

The interactive access to the RCOM objects is provided {\it via} the main
OGLE WWW page:

\begin{center}  
{\it http://ogle.astrouw.edu.pl}\\
\end{center}

The structure of the page and provided information are identical as for the
XROM system.

The photometry can be download from the OGLE archive:

\begin{center}  
{\it ftp://ftp.astrouw.edu.pl/ogle3/rcom}
\end{center}

\Acknow{This paper was partially supported by the Polish MNiSW grant
N20303032/4275. We thank Drs. Malcolm Coe, Matthew Schurch and Peter
Cottrell for encouraging us to design and implement presented systems.}


\begin{references}
\refitem{Afonso, C. \etal}{2000}{\ApJ}{532}{340}
\refitem{Alcock, C., Allsman, R. A., Alves, D. R. \etal}{2001}{\ApJ}{554}{298}
\refitem{Beaulieu, J.-P. \etal}{2006}{Nature}{439}{437}
\refitem{Coe, M.J., Edge, W.R.T., Galache, J.L., and McBride, V.A.}{2005}{\MNRAS}{356}{502}
\refitem{Dong, S., Udalski, A., Gould, A. \etal}{2007}{\ApJ}{664}{862}
\refitem{Gaudi, B.S., Bennett, D.P., Udalski, A. \etal}{2008}{Science}{319}{927}
\refitem{Gould, A., Udalski, A., An, D. \etal}{2006}{\ApJ}{644}{L37} 
\refitem{McGovan, K.E., Coe, M.J., Schurch, M.P.E.; Corbet, R.H.D.,
Galache, J.L., and Udalski, A.}{2008}{\MNRAS}{384}{821}
\refitem{Tisserand, P. \etal}{2004}{\AA}{424}{245}
\refitem{Tisserand, P. \etal}{2008}{\AA}{481}{673}
\refitem{Udalski, A., Szymański, M.; Kałużny, J., Kubiak, M., Mateo, M.}{1992}{\Acta}{42}{253}
\refitem{Udalski, A., Szyma\'nski, M., Ka{\l}u{\.z}ny, J., Kubiak, M.,
Mateo, M., Krzemi{\'n}ski, W., and Paczy{\'n}ski, B.}{1994}{\Acta}{44}{227}
\refitem{Udalski, A., Kubiak, M., and Szyma\'nski, M.}{1997}{\Acta}{47}{319}
\refitem{Udalski, A.}{2003}{\Acta}{53}{291}
\refitem{Udalski, A.}{2004}{IAUC}{~}{8276}
\refitem{Udalski, A., Jaroszyński, M., Paczyński, B. \etal}{2005}{\ApJ}{628}{109}
\refitem{Udalski, A. \etal}{2006}{\Acta}{56}{293}
\refitem{Wo{\'z}niak, P.R., Alard, C., Udalski, A., Szymański, M.,
Kubiak. M., Pietrzyński, G., and Żebruń, K.}{2000}{\ApJ}{529}{88} 
\end{references}
\end{document}